\documentclass[]{spie}  %>>> use for US letter paper
\usepackage{amsmath,amsfonts,amssymb}
\usepackage{graphicx}
\usepackage[colorlinks=true, allcolors=blue]{hyperref}
\usepackage{enumitem}
\usepackage{booktabs}
\usepackage{placeins}
\usepackage{siunitx}
\usepackage[bottom]{footmisc}
\usepackage{authblk}

\title{Reproducibility and monitoring of the instrumental particle background for the X-Ray Integral Field Unit}

\author[a]{E. Cucchetti}
\author[a]{E. Pointecouteau}
\author[a]{D. Barret}
\author[b]{S. Lotti}
\author[b]{C. Macculi}
\author[d]{S. Molendi}
\author[a]{F. Pajot}
\author[c]{P. Peille}
\author[b]{L. Piro}
\author[e,f]{G.~W. Pratt}

\affil[a]{\normalsize IRAP, Universit\'{e} de Toulouse, CNRS, CNES, 9 Av. du Colonel Roche,  31400 Toulouse, France} \affil[b]{INAF, IAPS, Via Fosso del Cavaliere 100, 00133, Roma}
\affil[c]{CNES, 18 Avenue Edouard Belin 31400 Toulouse, France}
\affil[d]{INAF, IASF Milano, Via E. Bassini 15, I-20133 Milano, Italy}
\affil[e]{IRFU, CEA, Universit\'{e} Paris-Saclay, F-91191 Gif-sur-Yvette, France}
\affil[f]{Universit\'{e} Paris Diderot, AIM, Sorbonne Cit\'{e}, CEA, CNRS, F-91191 Gif-sur-Yvette, France}

\authorinfo{Further author information: (Send correspondence to Edoardo Cucchetti) \\  \hspace*{1.6em} Edoardo Cucchetti: E-mail: edoardo.cucchetti@irap.omp.eu}

\begin{document} 
\maketitle

\begin{abstract}
The X-ray Integral Field Unit (X-IFU) is the cryogenic imaging spectrometer on board the future X-ray observatory \textsl{Athena}. With a hexagonal array of 3840 AC-biased Transition Edge Sensors (TES), it will provide narrow-field observations (5' equivalent diameter) with unprecedented high spectral resolution (2.5~eV up to 7~keV) over the 0.2 -- 12~keV bandpass. Throughout its observations, the X-IFU will face various sources of X-ray background. Specifically, the so-called Non-X-ray Background (NXB) caused by the interaction of high-energy cosmic rays with the instrument, may lead to a degradation of its sensitivity in the observation of faint extended sources (e.g. galaxy clusters outskirts). To limit this effect, a cryogenic anti-coincidence detector (CryoAC) will be placed below the detector plane to lower the NXB level down to the required level of $5 \times 10^{-3}$~cts/s/cm$^{2}$/keV over 2 -- 10~keV. In this contribution, we investigate ways to accurately monitor the NXB and ensure the highest reproducibility in-flight. Using the limiting science case of the background-dominated observation of galaxy clusters outskirts, we demonstrate that a reproducibility of 2\% on the absolute knowledge of the background is required to perform driving science objectives, such as measuring abundances and turbulence in the outskirts. Monitoring of the NXB in-flight through closed observations, the detector's CryoAC or the companion instrument (Wide Field Imager) will be used to meet this requirement.
\end{abstract}


\section{Introduction}

The launch in the 2030s of the next generation X-ray observatory \textit{Athena} will provide a leap forward in X-ray astronomy. With its two instruments -- the X-ray Integral Field Unit (X-IFU) [\citen{Barret2016XIFU}] and the Wide Field Imager (WFI) [\citen{Rau2013WFI}] -- \textit{Athena} will address a large number of scientific questions related to the Hot and Energetic Universe, spanning from the study of the chemical enrichment of the Universe to the properties of compact objects such as X-ray binaries or Black Holes [\citen{Nandra2013Athena}]. With its spatial capabilities (5'' resolution over a 5' equivalent diameter field-of-view) coupled to an excellent spectral resolution (2.5~eV required up to 7~keV), the X-IFU will provide spatially-resolved high-resolution spectroscopy in the soft X-ray band (0.2~--~12~keV) to study the fine spectroscopic features of the X-ray sky (e.g. line shifts, line broadening).

The focal plane of the X-IFU will be composed of a hexagonal array of $\sim$ 3840 Transition Edge Sensors (TESs) [\citen{Smith2016Pix}] micro-calorimeters, cooled to superconducting temperatures of 90~mK and read out using a Frequency Domain Multiplexing (FDM) scheme [\citen{Akamatsu2018FDM,Ravera2014DRE}]. Micro-calorimeters detect the increase of temperature due to the thermalisation of an X-ray photon in their absorber, which causes a sharp change in their intrinsic resistance over a short duration in time. As the detectors are voltage biased in their superconducting transition, this change in resistance is seen as a current pulse, which can then be filtered [\citen{Moseley1988Opt}] to deduce the energy of the incident photon. However, any photon from astrophysical sources beside the source of interest and within the detector energy bandpass will cause additional pulses, which will contaminate the primary science information. This component is defined here as the X-ray background. If unaccounted for, this component can cause systematic errors in the data interpretation but also degradations in the line sensitivity. 

Part of this background can be attributed to the ``Non-X-ray background'' (NXB). This component is caused by the interaction of high-energy particles (cosmic rays) with the spacecraft/instrument, which will in turn generate low-energy secondary particles capable of interacting with the TES array. In this contribution, we give a specific look at the NXB. After presenting the various sources of background and our assumptions (Sect. \ref{sec:bkg}), the effects of the NXB on the science of the X-IFU are investigated. Starting from the driving science case of galaxy clusters (i.e., faint extended sources), the effect of an error in the knowledge of the level of NXB (i.e., its reproducibility) is assessed (Sect. \ref{sec:rep}). Consequently, various techniques to monitor this background in-flight are investigated (Sect. \ref{sec:mon}).

\section{Sources of background on the X-IFU}
\label{sec:bkg}
\begin{figure}[tbp]
\centering
\includegraphics[width=0.49\textwidth, clip=True, trim={0cm 0cm 0cm 0cm}]{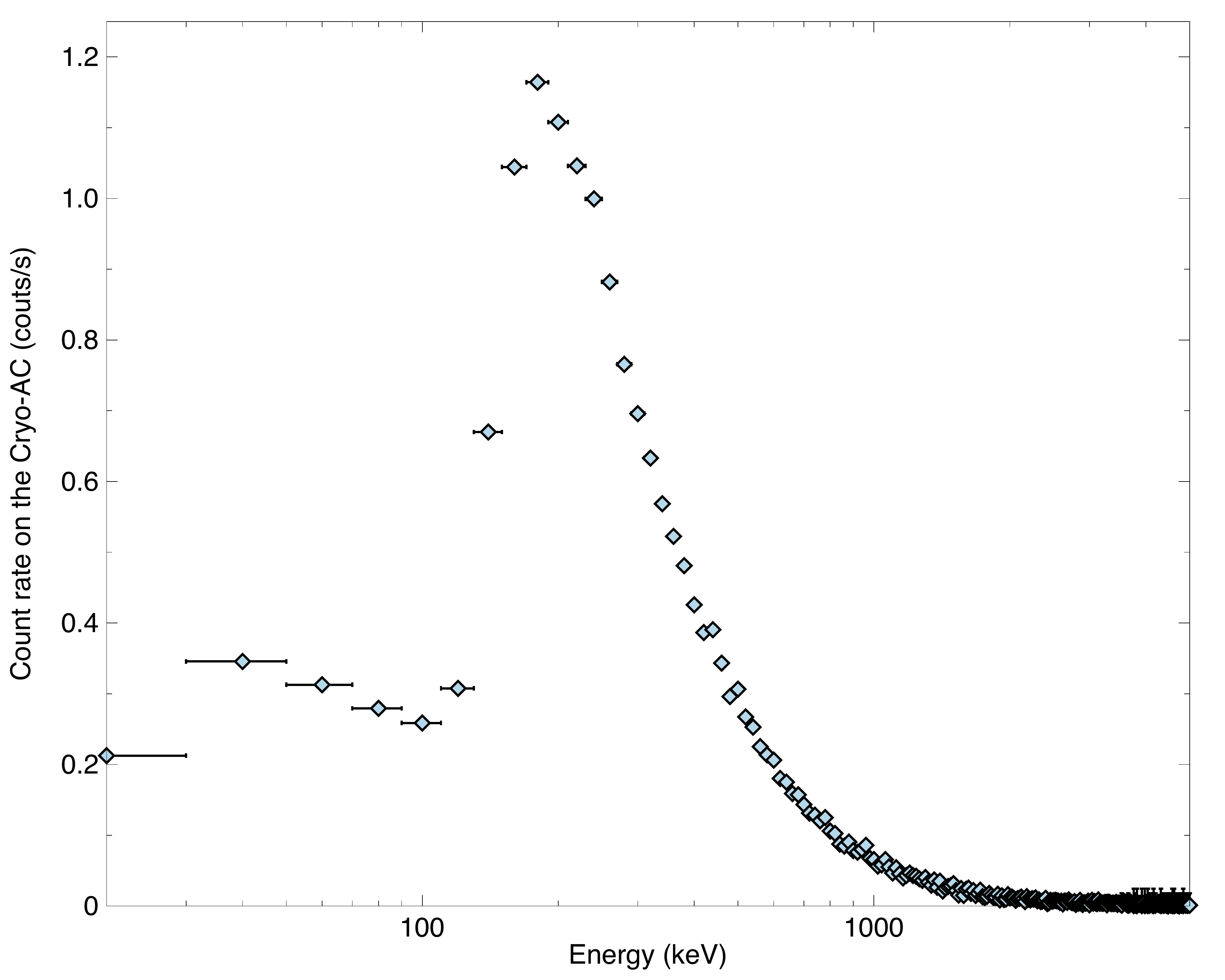}
\includegraphics[width=0.49\textwidth, clip=True, trim={0cm 0cm 0cm 0cm}]{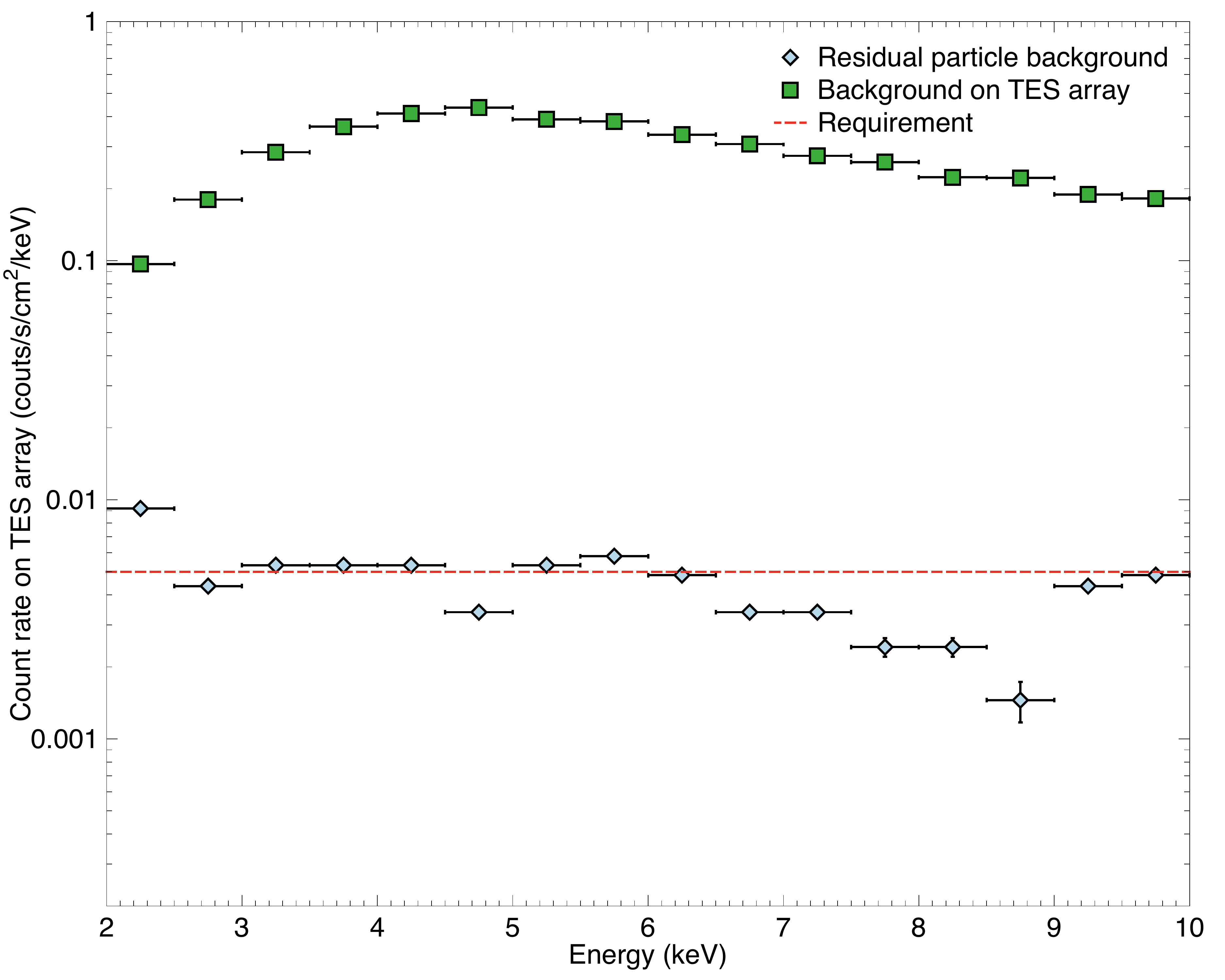}
\caption{(\textit{Left}) Count rate (cts s$^{-1}$) of the total Galactic Cosmic Background seen by the Cryo-AC over the 20 -- 800~keV energy band simulated using \texttt{GEANT-4} for $\sim$~900~s. The low-energy peak ($\sim$~40~keV) is due to secondary particles, while primaries are responsible for the higher energy peak ($\sim$~300~keV). (\textit{Right}) Count rate (cts s$^{-1}$ cm$^{-2}$ keV$^{-1}$) seen by the entire TES array before (green squares) and after the CryoAC background reduction (blue diamonds). The red dashed line indicates the requirement of $5 \times 10^{-3}$ cts s$^{-1}$ cm$^{-2}$ keV$^{-1}$.}
\label{fig:tes}
\end{figure}

The background expected for the X-IFU can be divided into multiple components:
\begin{itemize}
\item A first component, referred to as the ``astrophysical X-ray foreground'', is related to the local environment of the solar system (i.e., local bubble) and at larger scales to the halo of hot emitting gas in the Milky Way. It is well described by the sum of an unabsorbed and an absorbed diffuse thermal plasma emission model [\citen{McCammon2002Bkg, Lotti2014Bkg}]. 
\item The Sun will also significantly contribute to the instrumental background via a strongly variable component of emitted soft protons [\citen{DeLuca2004Soft}]. These protons will be focused by the optics onto the detector and create particles in the soft X-ray band visible by the TES array. A dedicated soft proton magnetic diverter will be placed in the spacecraft to deflect most of these protons away from the focal plane [\citen{Perinati2018Div}]. In this contribution, we assume \textbf{a perfect magnetic diverter}, i.e. the soft proton component is not considered in this study.
\item The sky is dotted with X-ray sources such as quasars, Active Galactic Nuclei (AGNs) or clusters, which define the ``Cosmic X-ray Background'' (CXB) [\citen{Lehmer2012AGN}]. In the case of the X-IFU, given the spatial resolution of the instrument, up to 80\% of the log(N)-log(S) integrated flux of these sources should be resolved [\citen{Moretti2003CXB}]. For typical 100~ks observations, point-like sources with fluxes $\gtrsim 3 \times 10^{-16}$ ergs/s/cm$^{2}$ will therefore be excised from observations. The unresolved part of the CXB however will result in a diffuse component in the observations. Although uncertain and likely spatially-dependent, the residual CXB is modelled here by an absorbed power-law component.
\item The final component of the X-ray background is called the internal particle background or Non-X-ray Background (NXB). High-energy primary particles (cosmic rays) will interact with the overall spacecraft to create showers of secondary particles, many of which fall in the soft X-ray band. The combination of these primary and secondary particles generates the NXB.
\end{itemize}

In the rest of the paper, a dedicated look is given to this last background component. Notably, we assume that the astrophysical foreground and the CXB are perfectly known. The spectrum of the NXB is closely related to that of the galactic cosmic rays (GCR), and will depend on the overall geometry of the spacecraft. Estimates of this component are found using \texttt{GEANT-4} simulations [\citen{Agostinelli2003GEANT}] with the most updated mass models of the instrument for the most conservative case of the GCR spectrum at solar minimum [\citen{Lotti2018BKG}]. The NXB will be lowered to a level of $5 \times 10^{-3}$~cts~s$^{-1}$~cm$^{-2}$~keV$^{-1}$ over the 2 -- 10 keV spectral bandpass using a cryogenic anti-coincidence detector (or CryoAC) [\citen{Macculi2016Cryo,D'Andrea2017CryoAC}]. Placed underneath the main array, detections on the CryoAC will `flag' background particles and reduce this component to the required level. Figure \ref{fig:tes} (\textit{Left}) shows the simulated level of background seen on the CryoAC, with a high-energy part related to primary particles (peak around $\sim$~300~keV) and a low-energy peak from secondary particles (around $\sim$~30 -- 40~keV). Figure \ref{fig:tes} (\textit{Right}) shows the simulated level of background on the TES without CryoAC and the actual NXB level achieved after the CryoAC background reduction. 

\section{Reproducibility of the NXB}
\label{sec:rep}

The NXB will dominate over the other background components at higher energies (typically $\geq$ 2~keV). It will thus mostly affect high-energy observations of faint sources, for which the count rate is comparable to the level of background. Its variability and the small collective area of the detector ($\sim$~19.5~arcmin$^{2}$) make this component difficult to estimate in practice. Yet, small errors in the knowledge of the NXB may cause significant systematic effects on the science results. The knowledge of the level of NXB (i.e. the maximal error on its level) is defined as its reproducibility. In this section, we investigate the reproducibility of the NXB over a solid angle of 9~arcmin$^{2}$ (as per the current \textit{Athena} requirements) by taking as driving science case the observation of galaxy clusters from their center to their outskirts (defined here in terms of the scale radius $R_{500}$\footnote{$R_{500}$ is the radius encompassing an overdensity of 500 times the critical density of the Universe at the given redshift.}).

\subsection{Simulation setup}

The hot gas emission of a `toy' model cluster is simulated by the product of a thermal plasma model (\texttt{bapec} [\citen{Smith2001APEC}] on XSPEC [\citen{Arnaud1996XSPEC}]) and of a \texttt{wabs} absorption model [\citen{MorrisonWabs}], with a column density $n_{\text{H}}=0.03\times 10^{22}$~cm$^{-3}$, a turbulent velocity of $V=500$~km/s, a constant metal abundance of Z=0.3~Z$_{\odot}$ (as per [\citen{Anders1989Solar}]). The cluster is taken at a local redshift of $z=$~0.1 with a central temperature of $kT=8$~keV. A universal emission measure (EM) profile is derived from [\citen{Eckert2012Emission}] and a temperature profile from [\citen{Reiprich2013Outskirts}] is assumed. We also considered a self-similar evolution of the clusters, spherical symmetry and scaling relations between the temperature and the scaled radius for massive clusters taken from [\citen{Arnaud2005Scaling}]. With these inputs, the fiducial scale radius $R_{500}$ is 1.4~Mpc, i.e., covering an angle of $\theta_{500}$= 12.1~arcmin on the sky.\footnote{where a $\Lambda$-CDM cosmology with $h = 0.7$, $\Omega_{m} = 0.3$ and $\Omega_{\Lambda} = 0.7$ is assumed}

\begin{figure}[tb]
\centering
\includegraphics[height=0.35\textwidth, clip=True, trim={0cm 1.5cm 3cm 1.5cm}]{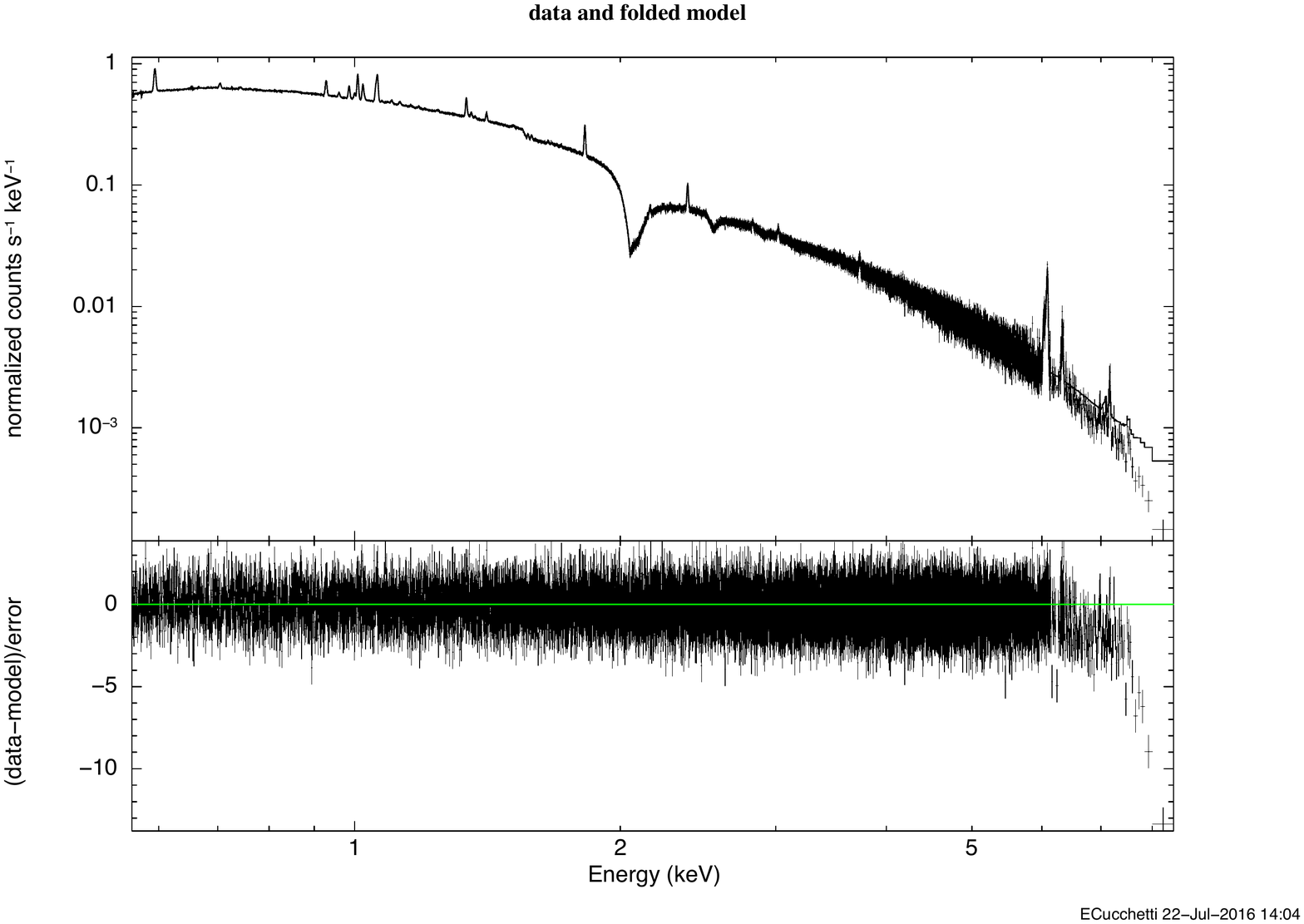}
\includegraphics[width=0.49\textwidth, clip=True, trim={0cm 0cm 0cm 0cm}]{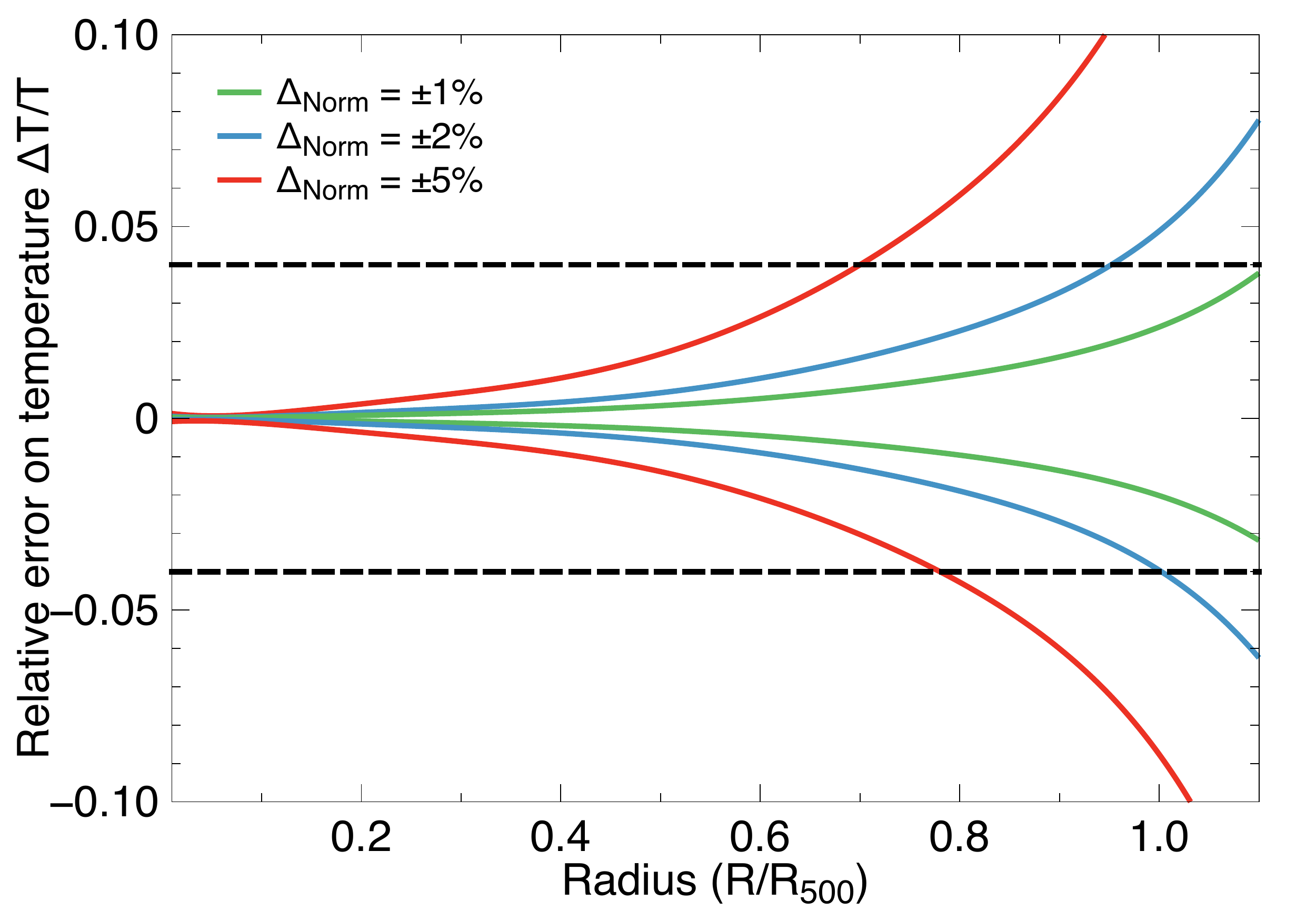}
\includegraphics[width=0.49\textwidth, clip=True, trim={0cm 0cm 0cm 0cm}]{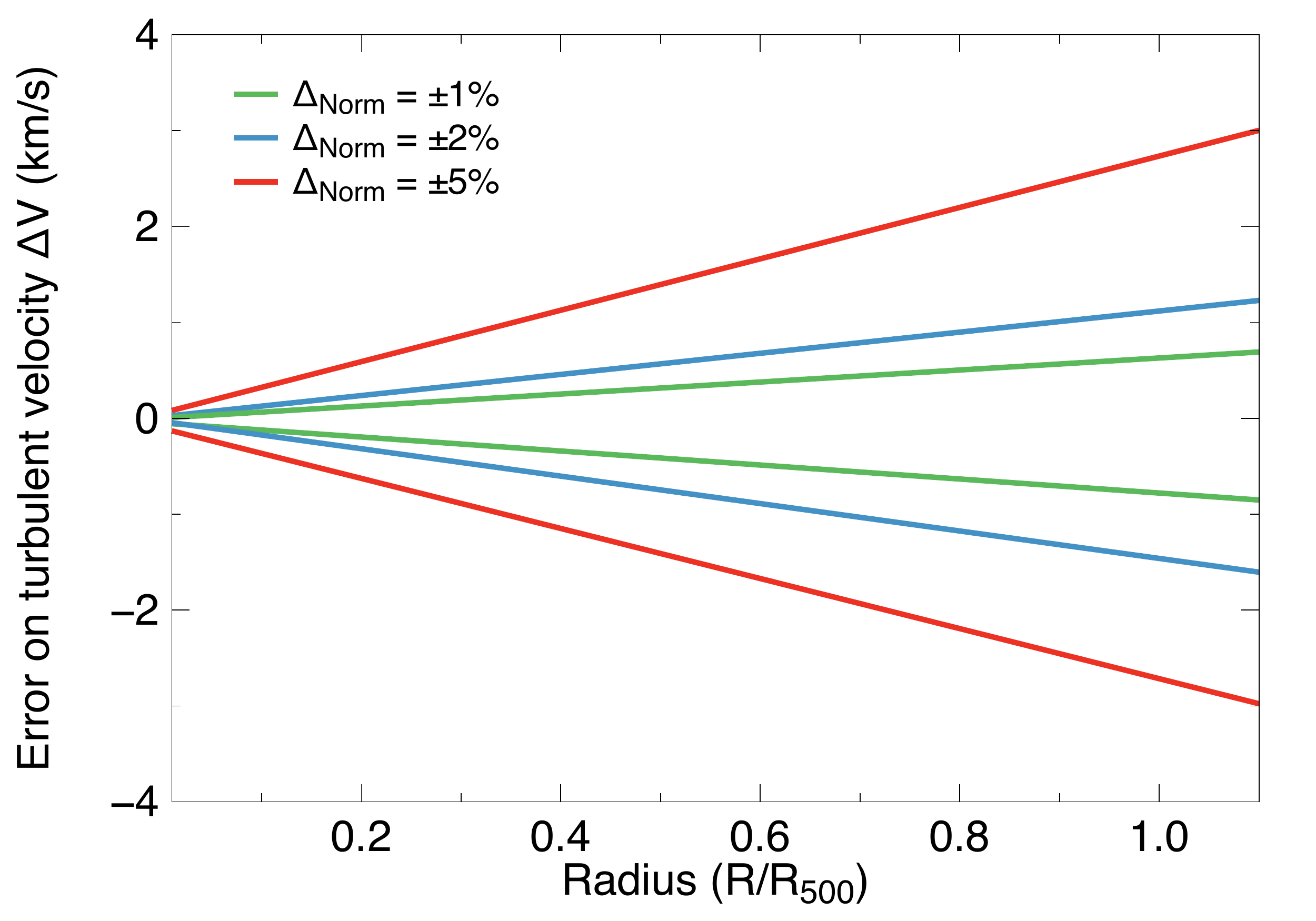}
\hspace{-0.4cm}
\includegraphics[width=0.49\textwidth, clip=True, trim={0cm 0cm 0cm 0cm}]{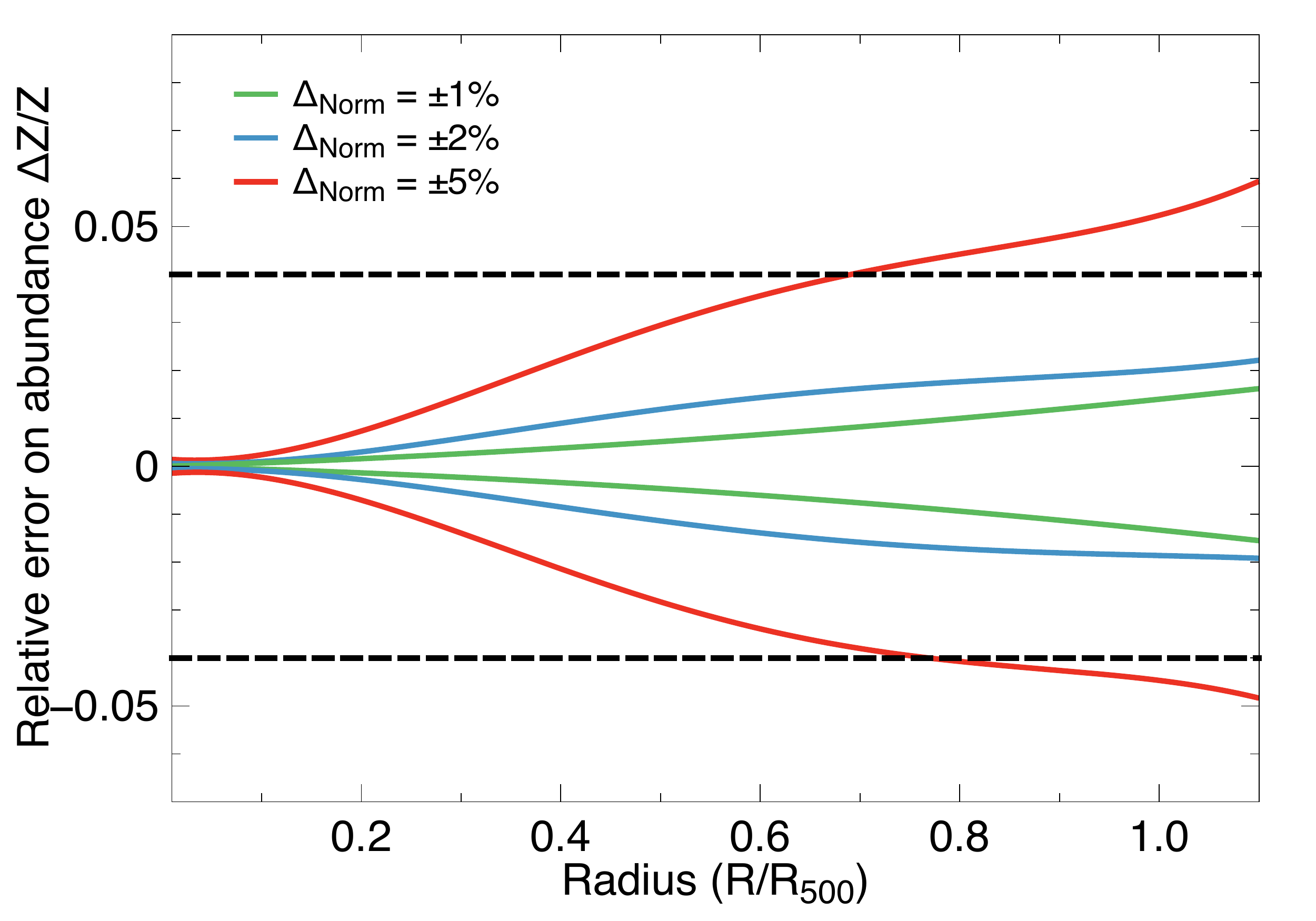}
\caption{(\textit{Top left}) Example of a spectrum fitted with a biased estimation of the NXB normalisation by $\Delta_{\text{Norm}}$=2\% for an unrealistic 100~Ms observation (to observe systematic effects). Residuals show an underestimation at high energy related to an over-subtraction of the background. Corresponding systematic errors made over the radial profile of the cluster (in units of $R_{500}$) for different values of  $\Delta_{\text{Norm}}=\pm$ 1 (green), 2 (blue) and 5\% (red) for temperature (\textit{Top right} - relative error), abundances (\textit{Bottom right} - relative error) and turbulent velocity (\textit{Bottom left} - in km/s). Dashed lines indicate the current 4\% requirement accuracy for temperature and abundances.}
\label{fig:syste}
\end{figure}

To assess the reproducibility of the NXB, the cluster is divided into annuli, each of different temperature and emission measure, constant over the considered 9~arcmin$^{2}$. The spectrum in each annulus is simulated using XSPEC by adding a constant NXB component corresponding to the requirement along with a sum of the other astrophysical components (foreground and CXB, assumed to follow the models given in [\citen{Lotti2014Bkg}] and references therein). Two sets of simulations were conducted. First this spectrum is fitted after a perfect background subtraction to estimate the level of statistical reproducibility (simulations `A'). This step determines the statistical error $\sigma_{\text{stat}}(R, t_{\text{exp}})$ of a given parameter as a function the radius $R$, for an exposure time $t_{\text{exp}}$. The same spectrum is also fitted after introducing a wrong estimation on the NXB normalisation $\Delta_{\text{Norm}}$ during the background subtraction, to evaluate the systematic effects due to the reproducibility (simulations `B'). Spectra are fitted over the entire energy band after binning using C-statistics [\citen{Cash1979}] leaving as free parameters the normalisation, temperature, abundances and turbulent velocity (see Figure \ref{fig:syste} -- \textit{Top left} for an example). This procedure was repeated 100 times, with fixed systematic shifts of $\Delta_{\text{Norm}}=\pm$ 1, 2 and 5\% (see also \citen{Molendi2016Background} for a similar approach).

\subsection{Systematic effect}

A first investigation on the reproducibility is carried out by considering exclusively systematic effects. To reduce as much as possible the statistical error (Poisson noise) due to the lack of counts in regions of faint emission and to assess solely the systematic error induced by the variation of the NXB normalisation, all the spectra were simulated for the (very unrealistic) exposure time of 100~Ms. In this configuration, we find using the dataset from simulations `A' that for each of the free parameters under consideration (temperature, abundances and turbulent velocity), statistical errors are $\ll 0.1\%$, and therefore negligible with respect to systematics. The data from simulations `B' allow in turn to determine the corresponding error on the parameters as a function of the radius (in unit of $R_{500}$) for various $\Delta_{\text{Norm}}$ (Figure \ref{fig:syste}).

As expected, the systematic error on the parameters increases with radius as the emission measure decreases. The impact of a systematic uncertainty in the NXB has much higher impacts on the recovery of temperature and of abundances (for which the current considered requirement of a knowledge within 4\% is reached before $R_{500}$ for $\Delta_{\text{Norm}}=\pm$ 5\%) than on turbulent velocities, for which the systematic error is below 5~km/s over the considered radial extension, well below the required 20~km/s. This can be explained by the flat shape of the NXB. An error in the knowledge of the norm results mainly in a wrong estimation of the overall continuum value of the spectrum. This causes errors preferentially in temperature (related to the shape of the continuum) and abundances (integration of lines over the continuum) rather than on the broadening, which is relatively unaffected by a change in normalisation. Assuming a spatial scale of $R_{500}$, a reproducibility of 2\% is needed to reduce systematic errors below the corresponding threshold values for each of the parameters.
\begin{table}[!t]
\begin{center}
\caption{Impact of an error in the absolute knowledge of the NXB on the recovery of the temperature, $kT$, abundance, $Z$, and the turbulent velocity, $V$ for realistic exposure times of 100ks and 1Ms. Contribution of statistics are included.} .\\[0.1em]
\begin{tabular}{ S | S | SSSS  | SSSS } \toprule
&$t_{exp}$ &\multicolumn{4}{c|}{100 ~ksec} & \multicolumn{4}{c}{1~Ms} \\
 & $|\Delta_{\text{Norm}}|$ & 0\% &1\% &2\% &5\% & 0\% &1\% &2\% &5\% \\
 \midrule
  $kT$& $R_{\text{min}}\,^{\dag}$  &0.87& 0.87 &0.86 & 0.83 &1.5 & 1.43 & 1.31 & 1.0 \\
  & $\Delta_{\text{S}}\;\;^{\ddag}$  & {--} &0.11 & 0.2 & 0.7 & {--} & 0.43 & 0.99 & 2.4  \\
  & $\Delta_{\text{T}}\;\;^{\S}$  & {--} &0.12 & 0.27 & 0.57 & {--} &  0.41 & 0.67 & 0.93 \\
\midrule
  $Z$& $R_{\text{min}}$  &0.5 & 0.5 &0.49 & 0.45 & 1.10 & 1.09 & 1.07 & 0.72 \\
  & $\Delta_{\text{S}}$  & {--} &0.07 & 0.17 & 0.4 & {--} & 0.2 & 0.5 & 1.2 \\
  & $\Delta_{\text{T}}$  & {--} &0.07 & 0.15 & 0.4 & {--} & 0.2 & 0.45 & 0.8 \\
 \midrule
$V$& $R_{\text{min}}$  &0.22 & 0.22 &0.22 & 0.22 &0.61 & 0.61 & 0.61 & 0.61 \\
  & $\Delta_{\text{S}}$  & {--} &0.0015 & 0.003 & 0.01 & {--} & 0.015 & 0.02 & 0.05 \\
  & $\Delta_{\text{T}}$  & {--} &0.0015 & 0.003 & 0.01 & {--} & 0.01 & 0.02 & 0.06 \\
\bottomrule
\end{tabular}
\label{tab:stat}
\end{center}
\footnotesize
$^{\dag}$ $R_{\text{min}}$ is the minimal radius at which the error on the parameter reaches the requirement in units of $R_{500}$. \\
$^{\ddag}$ $\Delta_{\text{S}}$ is the ratio of  the systematic to the statistical error. The maximal value of  $\sigma_{\text{syst}}$  for $\pm$ 1, 2 and 5 \% is used.\\
$^{\S}$ $\Delta_{\text{T}}$ is the ratio of  the systematic to the total error, i.e., the  systematic error contribution to the total error budget.
\normalsize
\end{table}

\subsection{Influence of statistics}

Although systematic effects may have strong impacts on the observation if unaccounted for, previous simulations used very high (unrealistic) exposure times, therefore giving no information on the scale of the statistical error with respect to systematics in the case of more realistic exposure times. To do so, the previous simulation setup was used for 100~ks and 1~Ms exposures, representing respectively typical observations and `deep' pointings. Results are summarised in Table \ref{tab:stat} along with the corresponding ratios between statistical to systematic error and of systematic to total error. We also verified that systematic errors were consistent with results found for very high exposure times.  The low accuracy on some of the parameters is related to the choice of the observed region (9 arcmin$^{2}$), in accordance to the corresponding \textit{Athena} requirement. For real observations of faint regions, larger regions need to be covered to ensure an accurate result.

Statistical errors dominate the error budget for all the physical parameters. When exposure time increases, both errors become comparable, although the statistical error remains dominant at large radii, due to lower count rates in the cluster outskirts. Using these results, we define a suitable level of reproducibility for the NXB which satisfies, for 100~ks observations:
\begin{enumerate}
\itemsep0em
\item A systematic error below requirements up to $R_{500}$
\item A level of systematic errors small with respect to the statistical error (i.e. $5 \sigma_{\text{syst}} \leq \sigma_{\text{stat}}$) up to $R_{500}$
\end{enumerate}
\noindent Under this assumption, the level of reproducibility is taken in the rest of this paper at 2\% of the nominal value of the NXB. Whenever larger exposure times are considered condition 2 can be less restrictive since statistical and systematic error become comparable. In which case, the requirement can be relaxed up to $\sim$ 3\%, when condition 1) is no longer met.

\section{Monitoring the NXB}
\label{sec:mon}

We discuss in this section the different possibilities at our disposal to monitor the NXB in-flight to monitor its level with an accuracy better than 2\%. 

\subsection{Closed observations with the X-IFU}
\label{subsec:closed}
\begin{figure}[tb]
\centering
\includegraphics[width=0.49\textwidth, clip=True, trim={0cm 0cm 0cm 0cm}]{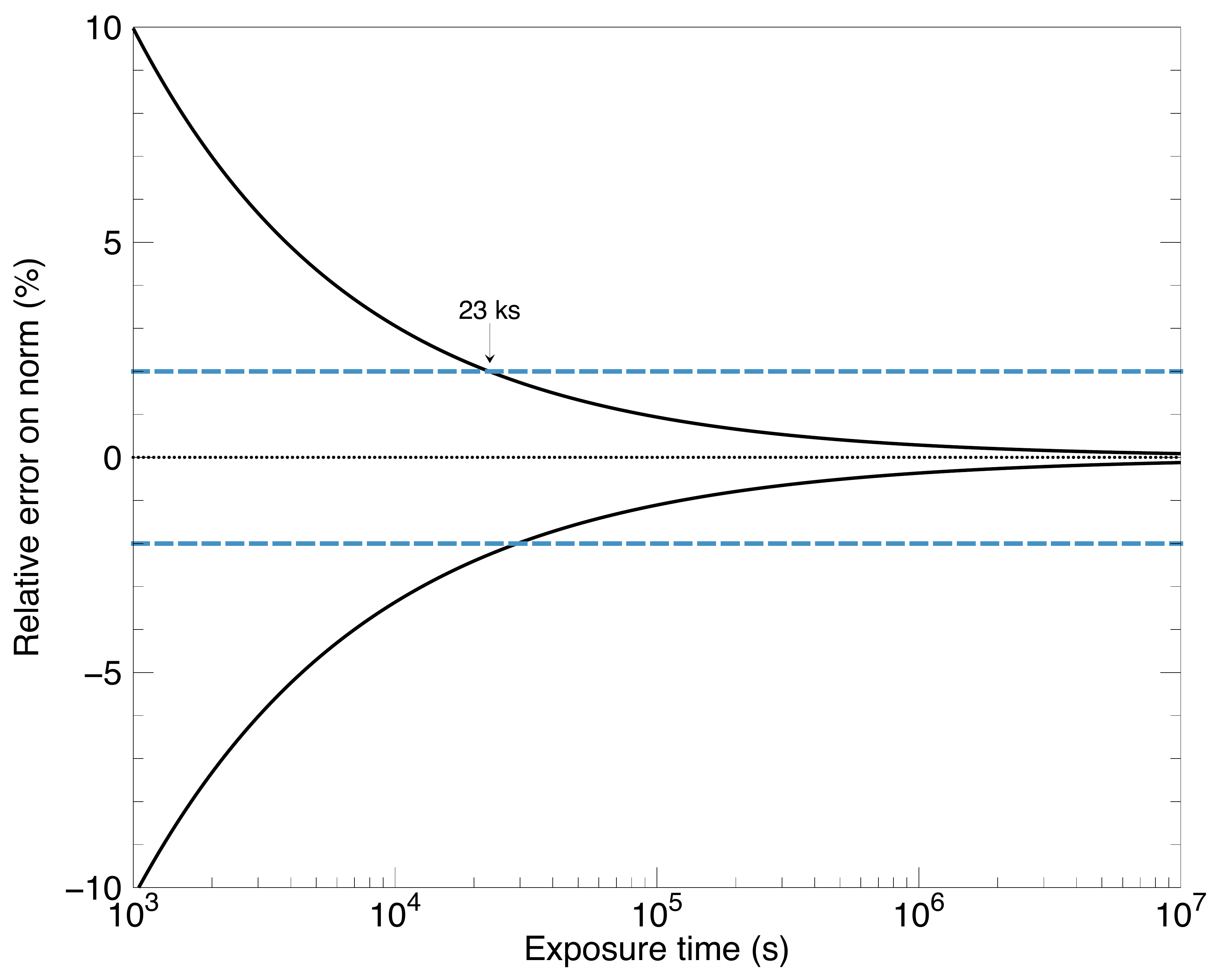}
\includegraphics[width=0.49\textwidth, clip=True, trim={0cm 0cm 0cm 0cm}]{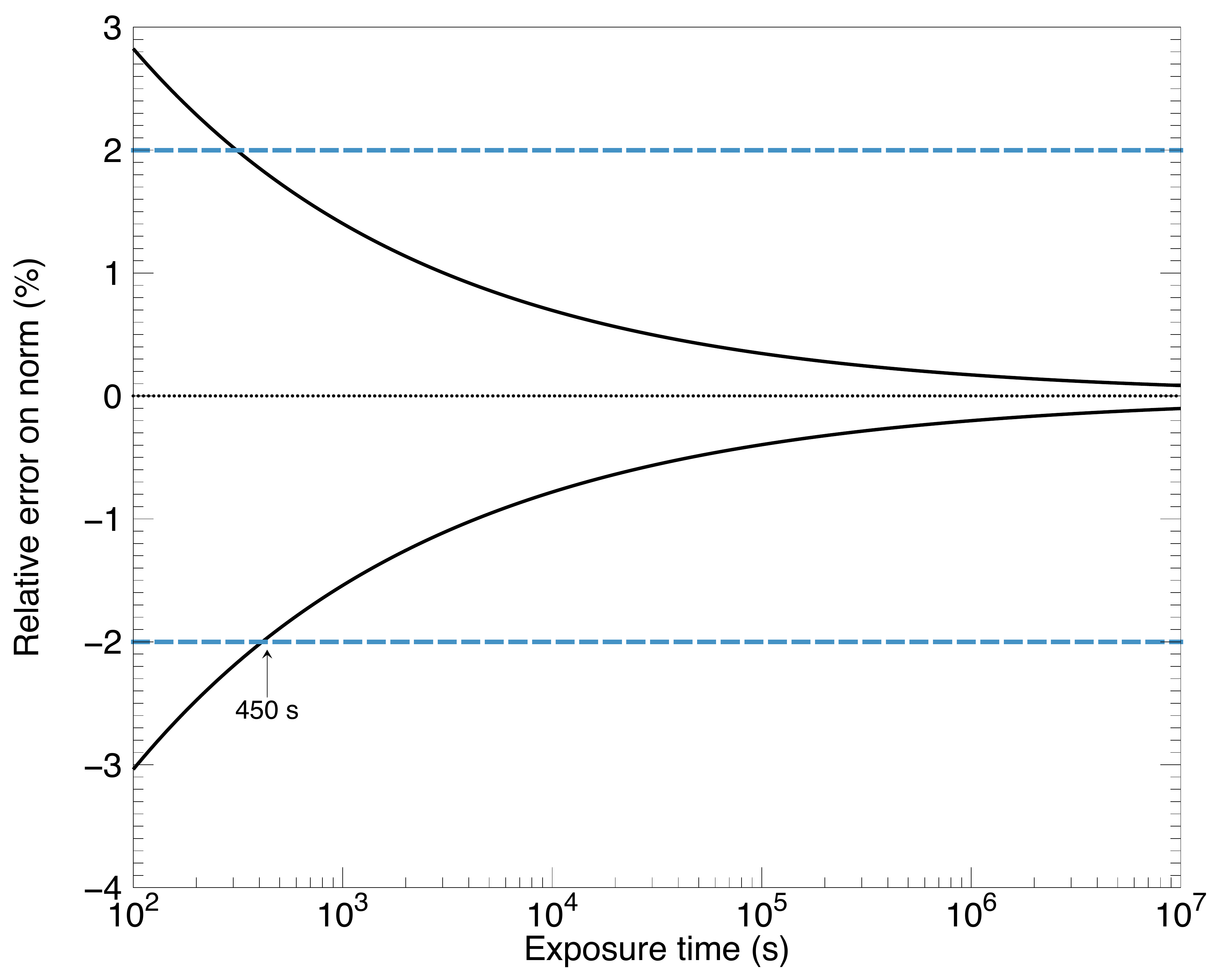}
\caption{(\textit{Left}) Relative error (\%) on the value of the NXB normalisation ($\pm 1 \sigma$) as a function of time (s) used for a closed observation of the X-IFU. (\textit{Right}) Likewise for the WFI assuming the same level of NXB.  Dashed lines represent the 2\% threshold. }
\label{fig:closed}
\end{figure}

The simplest solution to monitor the NXB for the X-IFU is to use closed observations of the instrument (i.e., the filter wheel of the instrument in `closed' position [\citen{Bozzo2016FW}]), during which the only events seen by the TES array can be attributed to the NXB. To assess the performances of these observations, a constant level of NXB was simulated for different closed observation times. The corresponding error on the normalisation as a function of time can be obtained by repeating the process a large number of times (Figure \ref{fig:closed} -- \textit{Left}).

Despite its simplicity, this solution presents several drawbacks. First of, since the field-of-view of the X-IFU is rather small ($\sim$~19.5~arcmin$^{2}$), a significant exposure time is needed to decrease the uncertainty on the knowledge of the NXB level below 2\%. As shown Figure \ref{fig:closed} (\textit{Left}), $\sim 20$~ks  are needed to know this level at better than $1\sigma$, which correspond to $\sim$ 3000 counts in the bandpass for the considered flat NXB spectrum. Further, during observations, the astrophysical data is collected assuming the same level of NXB as the one estimated during closed periods. Although the variability of the GCR should be low in time (11 year solar cycle), slight changes in overall normalisation may occur during $\geq$ 100~ks observations. The difference in the level can be accounted for by renormalising the value obtained during closed observations using the high-energy band (e.g., $\geq$~10~keV). However, this implies that no source emission is present and that no variation in the slope of the NXB occurs.

\subsection{Using the CryoAC}

During observations, the CryoAC will be used to monitor and flag background events on the array. Unlike the TES array, with its large area and its sensitivity at higher energies ($\geq 20$~keV), the CryoAC will accumulate a large number of counts over short time periods ($\sim 40$~cts/s over the entire detector). We investigate here -- purely as a feasibility study -- the idea of using these CryoAC counts as a way to monitor the particle background seen by the X-IFU. In the rest of this study we assume a configuration of the CryoAC which provides measurements over the 20 -- 800~keV bandpass. Though not required in the current baseline of the instrument, the CryoAC may possess spectral capabilities across the bandpass (as also investigated in [\citen{D'Andrea2017CryoAC}]). To perform this exercise, we assume the CryoAC will have such capabilities, with a guess estimate of the spectral resolution of 2~keV over the bandpass, sampled every 0.4~keV. Although such value will be unlikely over the entire bandpass, it is taken as starting point for simplicity. Future investigations are planned for other values of the energy resolution. Upcoming real-life tests will also provide more suitable energy dependent values of the energy resolution. For the NXB spectrum, a sampling in 0.4~eV steps (as in the current X-IFU responses) is considered. 

To exploit the CryoAC counts for background monitoring, the first step is to determine a link between the CryoAC and the residual background spectrum. A straightforward proportional relationship exists between the number of primaries seen by the CryoAC and the primaries seen by the TES array. However, assuming an isotropic distribution of the primaries, such relationship should also exist with the residual primaries on the detector. The same is considered for secondary particles. We therefore assume that the CryoAC spectrum $f_{\text{CryoAC}}$ (Figure \ref{fig:tes} -- \textit{Left}) and the residual particle background $f_{\text{NXB}}$ (Figure \ref{fig:tes} -- \textit{Right}) can be linked using a certain transfer function $H$. By construction, this function will relate the shape of the CryoAC spectrum to the residual background seen by the TESs, such that, for a given exposure time $t$ used for the observations we have:
\begin{align}
f_{\text{NXB}} (t) &= H \times f_{\text{CryoAC}}(t)
\label{eq:1}
\end{align}

The spectral response of the CryoAC is used here to estimate $f_{\text{CryoAC}}$. The main goal of this exercise is to determine the validity of this approach. In our case, as spectra measured by the instrument are discrete (with sizes corresponding to the number of spectral channels of the instrument), $H$ will be a rectangular matrix and its accuracy will depend on the accumulated time $t_{\text{acc}}$ used to compute it. Since $f_{\text{NXB}}$ and $f_{\text{CryoAC}}$ only take positive values, $H$ can be estimated by:
\begin{align}
H(t_{\text{acc}}) &= \frac{f_{\text{NXB}} (t_{\text{acc}}) \times f_{\text{CryoAC}}(t_{\text{acc}})^{T}}{|| f_{\text{CryoAC}}(t_{\text{acc}})||^{2}}
\label{eq:2}
\end{align}

\begin{figure}[tb]
\centering
\includegraphics[width=0.49\textwidth, clip=True, trim={0cm 0cm 0cm 0cm}]{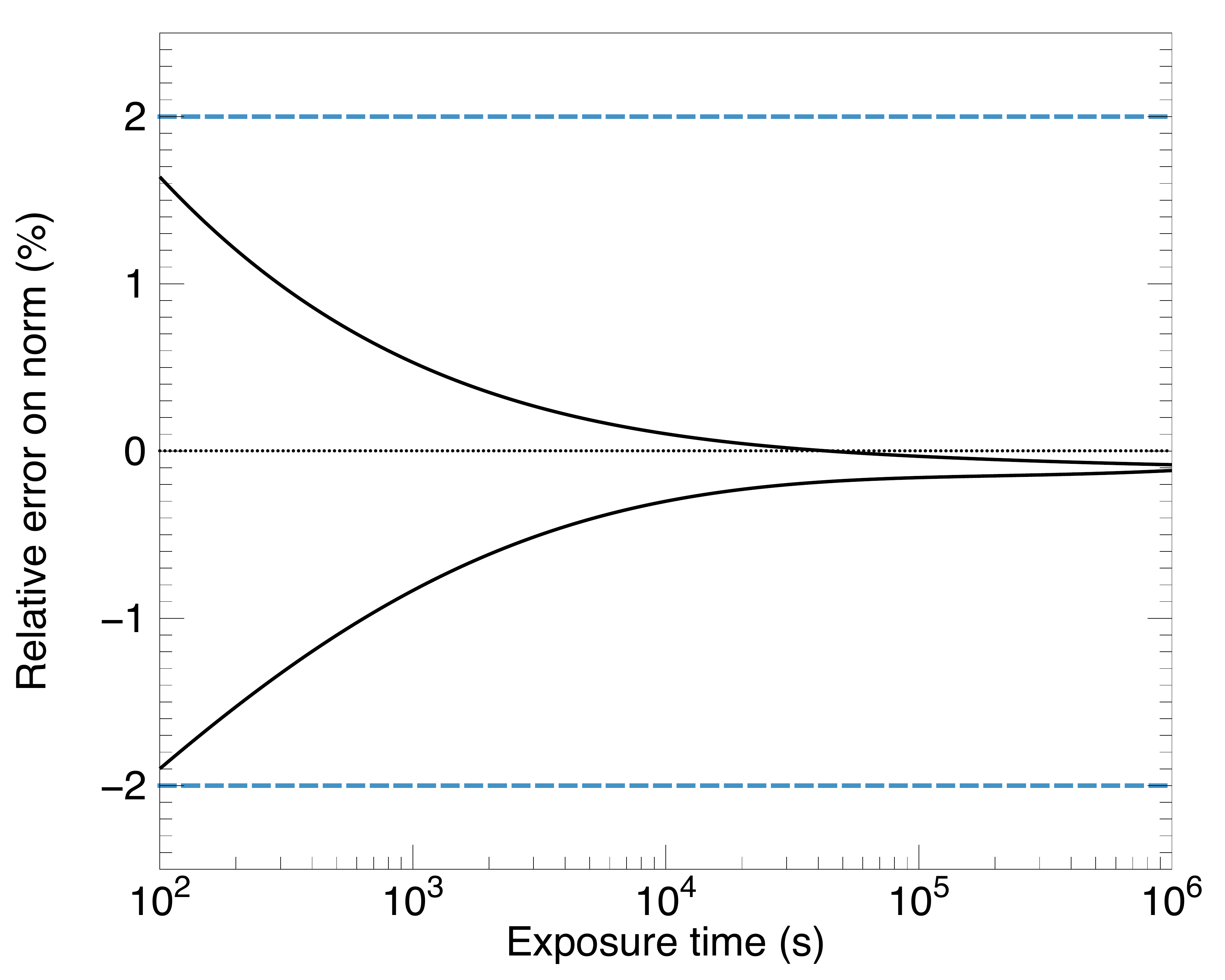}
\includegraphics[width=0.49\textwidth, clip=True, trim={0cm 0cm 0cm 0cm}]{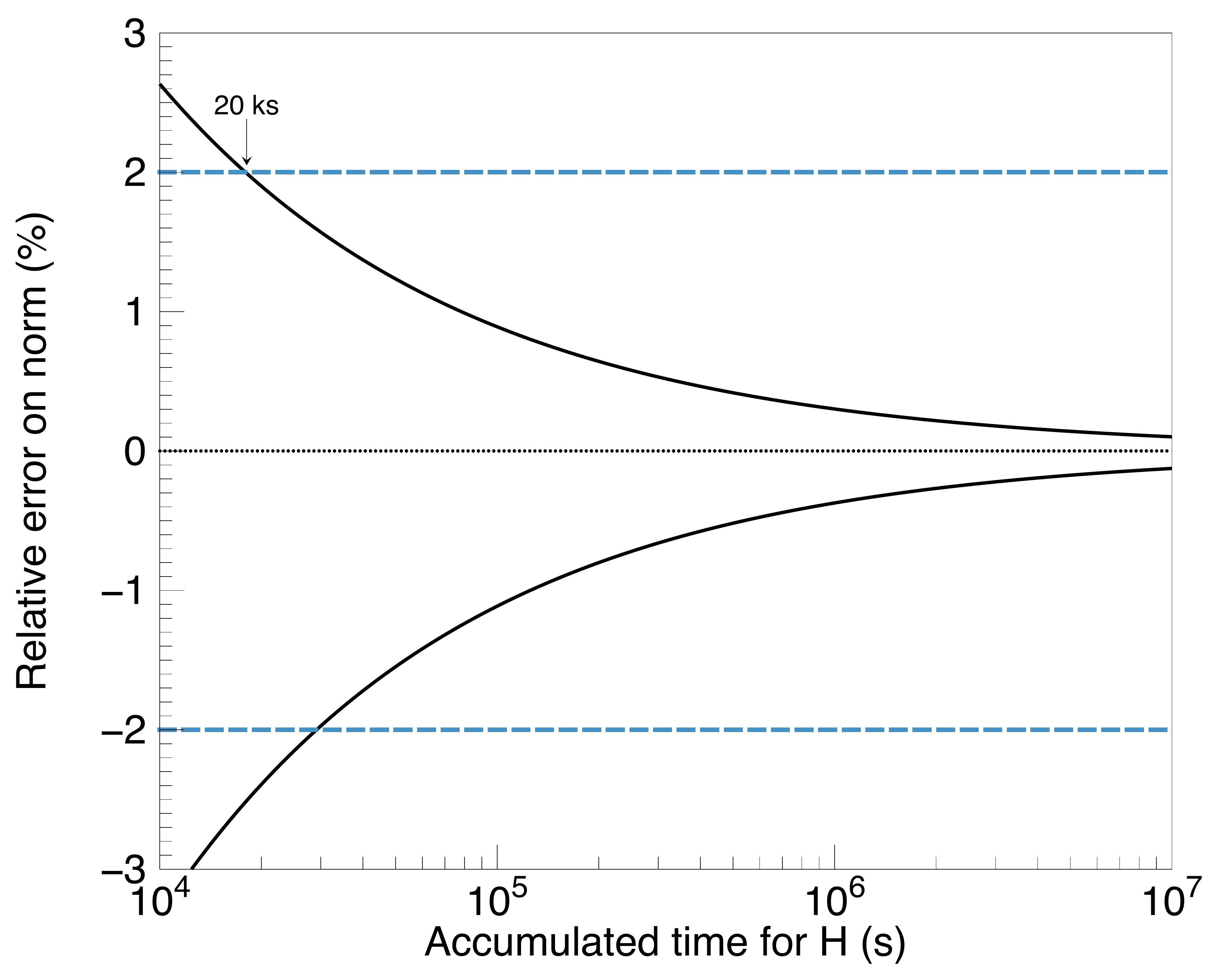}
\caption{(\textit{Left}) Relative error (\%) on the value of the NXB normalisation as a function of time (s) when using the CryoAC technique with an estimate of $H$ measured over 100~ks obtained for 1\,000 iterations. The bias on the correction is due to the finite accumulated time to estimate of $H$. (\textit{Right}) Relative error made on the CryoAC correction as a function of accumulated time (in s) used to estimate the transfer function matrix $H$.  Dashed lines represent the 2\% threshold. }
\label{fig:cryo}
\end{figure}

Equation~\ref{eq:2} was tested by computing an estimation of the matrix $H$ for an accumulated time of $t_{\text{acc}}$=100~ks. The values of the NXB spectrum were drawn randomly using the required level of $5 \times 10^{-3}$ cts s$^{-1}$ cm$^{-2}$ keV$^{-1}$. The CryoAC counts were instead randomly sampled using a toy model of the CryoAC spectrum, generated from the outputs of \texttt{GEANT-4} simulations with the latest mass model of the instrument. The recovery of an accurate level of NXB was assessed by taking CryoAC counts over a period of time $t_{\text{exp}}$ and applying Equation~\ref{eq:1} using $H$(100~ks) and repeating the process a large number of times ($\geq$ 1000). As shown Figure~\ref{fig:cryo} (\textit{Left}) this method allows to have a very accurate knowledge of the NXB level even for very short observations. 

Although accurate for high accumulated exposure times, systematic effects in the correction can appear if the transfer function is estimated with short accumulated times (see notably the shift with respect to zero on Figure~\ref{fig:cryo} -- \textit{Left}). To quantify this bias, we repeated for a given value of $t_{\text{acc}}$ the same process used to test Equation~\ref{eq:2}. By applying Equation~\ref{eq:1} for very long exposure times $t_{\text{exp}}$, the statistical error becomes very small, giving access to the systematic error on this given run. This process is then performed multiple times ($\geq$~100) to estimate the average error introduced by $H(t_{\text{acc}})$ on the observations. The corresponding behaviour of this bias with $t_{\text{acc}}$ is finally derived by extending this previous analysis for different values of the accumulated time (Figure \ref{fig:cryo} -- \textit{Right}). To achieve a sufficiently accurate value of $H$, we find that an accumulated exposure time $t_{\text{acc}}$ of the order of 20~ks is required, which is comparable to the result found Sect.~\ref{subsec:closed}.  

Similarly to the closed observation technique presented above, the use of the CryoAC also requires dedicated closed phases to estimate $H$. However, this option presents several advantages with respect to its counterpart. First of all, once computed, the transfer function can be used in parallel with observations, thus giving access to short-time variability of the NXB. Although small errors may persist if $t_{\text{acc}}$ is not sufficiently high, the accuracy of the method can be continuously improved in-flight by adding the CryoAC counts in Equation~\ref{eq:1}. Further, as shown by Equation~\ref{eq:2}, this technique is robust to changes in the overall normalisation of the CryoAC spectrum (but not to sharp changes in its spectral shape), which are likely to occur in-flight. Additional studies are therefore needed to investigate whether closed observations during dedicated calibration time would be sufficient to mitigate short-time variations of the NXB and achieve the desired accuracy. 

This entire study relies on the hypothesis that the CryoAC will have some spectral capabilities over its bandpass. Being an anti-coincidence detector, no spectral resolution is currently required nor quantified for the CryoAC in the baseline of the instrument. Changes in the assumed spectral resolution will affect the accuracy of the correction and shall be assessed in future studies. It is however the presence or not of these capabilities that will ultimately determine the overall feasibility of this option. 

\subsection{Other ways of monitoring the NXB}

\subsubsection{Using the companion instrument - WFI}

 Closed observations with the X-IFU are limited due to the narrow detector area, which requires long exposure times to observe a large number of photons. Assuming the WFI will see a similar level of NXB, the wider area of the companion instrument (0.45 deg$^{2}$ i.e., more than 40 times larger than the X-IFU) would allow to collect as many photons in a much smaller time. For the same level of accuracy, $\leq$~500~s of closed WFI observation would be needed (Figure \ref{fig:closed} -- \textit{Right}). Although appealing, this option requires that the two instruments can be accurately cross-calibrated while in-flight and also assumes a simultaneous operation of the WFI during X-IFU operations, both of which remain to be confirmed.

\subsubsection{High-energy band} 

The X-IFU will provide observations between 0.2 -- 12~keV, but higher energy detections may still be possible due to its high-energy effective area. If the NXB can be monitored over this bandpass (TBC), a direct measurement of the NXB level would be possible. This technique offers the advantage of accounting for time scale variability in the observation and could be done without any additional loss of time. Given the reduced effective area of the detector above 12~keV however (effective count rate of 0.012 cts/s over the detector between 12 and 13~keV if background monitoring is extended to high energies), an exposure time of tens of ks would be required to lower the error below the required 2\%.

\subsubsection{Radiation monitor}

\textit{Athena} is planned to have on board a radiation monitor, which will measure the flux of the X-ray environment on the satellite's orbit. By giving access to the overall flux of cosmic rays (and not the full spectrum -- TBC), the radiation monitor will provide important information on the overall time variability of the GCR, which can then be coupled to other methods to estimate the changes in the level of NXB.

\section{Conclusion}

In this paper, we have investigated through numerical simulations the effects of the NXB on X-IFU observations. Notably, the reproducibility of the NXB and its monitoring were addressed. Starting from the driving case of background-dominated galaxy clusters, we have studied in a first part the effect on the recovery of physical properties whenever the level of internal background was misestimated. Assuming 100~ks observations over a 9~arcmin$^{2}$ solid angle up to $R_{500}$ (as presented in the current \textit{Athena} requirements), we have demonstrated that a systematic knowledge of the level at better than 2\% is required to have an accurate recovery of parameters such as the temperature and the abundances of the hot thermal plasma in clusters of galaxies, while being negligible with respect to statistical effects. To monitor these effects, several solutions and ideas were presented, including the use of closed observation time of the companion instrument (the WFI) or of the CryoAC counts to estimate the level of NXB. Despite the large number of solutions available, their feasibility, robustness to short-time variability effects ($\leq$~1~ks) of the particle background and to the presence of solar soft protons remains to be confirmed. Further investigations are therefore required to ensure an accurate reproducibility of the background during observations.

% References
\bibliography{NXB_paper} % bibliography data in report.bib

\begin{thebibliography}{10}

\bibitem{Barret2016XIFU}
{Barret}, D., {Lam Trong}, T., {den Herder}, J.-W., {Piro}, L., {Barcons}, X.,
  {Huovelin}, J., {Kelley}, R., {Mas-Hesse}, J.~M., {Mitsuda}, K., {Paltani},
  S., {Rauw}, G., {Ro{\.Z}anska}, A., {Wilms}, J., {Barbera}, M., {Bozzo}, E.,
  {Ceballos}, M.~T., {Charles}, I., {Decourchelle}, A., {den Hartog}, R.,
  {Duval}, J.-M., {Fiore}, F., {Gatti}, F., {Goldwurm}, A., {Jackson}, B.,
  {Jonker}, P., {Kilbourne}, C., {Macculi}, C., {Mendez}, M., {Molendi}, S.,
  {Orleanski}, P., {Pajot}, F., {Pointecouteau}, E., {Porter}, F., {Pratt},
  G.~W., {Pr{\^e}le}, D., {Ravera}, L., {Renotte}, E., {Schaye}, J.,
  {Shinozaki}, K., {Valenziano}, L., {Vink}, J., {Webb}, N., {Yamasaki}, N.,
  {Delcelier-Douchin}, F., {Le Du}, M., {Mesnager}, J.-M., {Pradines}, A.,
  {Branduardi-Raymont}, G., {Dadina}, M., {Finoguenov}, A., {Fukazawa}, Y.,
  {Janiuk}, A., {Miller}, J., {Naz{\'e}}, Y., {Nicastro}, F., {Sciortino}, S.,
  {Torrejon}, J.~M., {Geoffray}, H., {Hernandez}, I., {Luno}, L., {Peille}, P.,
  {Andr{\'e}}, J., {Daniel}, C., {Etcheverry}, C., {Gloaguen}, E., {Hassin},
  J., {Hervet}, G., {Maussang}, I., {Moueza}, J., {Paillet}, A., {Vella}, B.,
  {Campos Garrido}, G., {Damery}, J.-C., {Panem}, C., {Panh}, J., {Bandler},
  S., {Biffi}, J.-M., {Boyce}, K., {Cl{\'e}net}, A., {DiPirro}, M., {Jamotton},
  P., {Lotti}, S., {Schwander}, D., {Smith}, S., {van Leeuwen}, B.-J., {van
  Weers}, H., {Brand}, T., {Cobo}, B., {Dauser}, T., {de Plaa}, J., and
  {Cucchetti}, E., ``{The Athena X-ray Integral Field Unit (X-IFU)},'' in [{\em
  Space Telescopes and Instrumentation 2016: Ultraviolet to Gamma
  Ray}{\nolinebreak\hspace{0.1em}]},  {\em Proc. SPIE} {\bf 9905},  99052F
  (2016).

\bibitem{Rau2013WFI}
{Rau}, A., {Meidinger}, N., {Nandra}, K., {Porro}, M., {Barret}, D.,
  {Santangelo}, A., {Schmid}, C., {Struder}, L., {Tenzer}, C., {Wilms}, J.,
  {Amoros}, C., {Andritschke}, R., {Aschauer}, F., {Bahr}, A., {Gunther}, B.,
  {Furmetz}, M., {Ott}, B., {Perinati}, E., {Rambaud}, D., {Reiffers}, J.,
  {Treis}, J., {von Kienlin}, A., and {Weidenspointner}, G., ``{The Hot and
  Energetic Universe: The Wide Field Imager (WFI) for Athena+},'' {\em ArXiv
  e-prints}  (Aug. 2013).

\bibitem{Nandra2013Athena}
{Nandra}, K., {Barret}, D., {Barcons}, X., {Fabian}, A., {den Herder}, J.-W.,
  {Piro}, L., {Watson}, M., {Adami}, C., {Aird}, J., {Afonso}, J.~M., and
  et~al., ``{The Hot and Energetic Universe: A White Paper presenting the
  science theme motivating the Athena+ mission},'' {\em ArXiv e-prints}  (June
  2013).

\bibitem{Smith2016Pix}
{Smith}, S.~J., {Adams}, J.~S., {Bandler}, S.~R., {Betancourt-Martinez}, G.~L.,
  {Chervenak}, J.~A., {Chiao}, M.~P., {Eckart}, M.~E., {Finkbeiner}, F.~M.,
  {Kelley}, R.~L., {Kilbourne}, C.~A., {Miniussi}, A.~R., {Porter}, F.~S.,
  {Sadleir}, J.~E., {Sakai}, K., {Wakeham}, N.~A., {Wassell}, E.~J., {Yoon},
  W., {Bennett}, D.~A., {Doriese}, W.~B., {Fowler}, J.~W., {Hilton}, G.~C.,
  {Morgan}, K.~M., {Pappas}, C.~G., {Reintsema}, C.~N., {Swetz}, D.~S.,
  {Ullom}, J.~N., {Irwin}, K.~D., {Akamatsu}, H., {Gottardi}, L., {den Hartog},
  R., {Jackson}, B.~D., {van der Kuur}, J., {Barret}, D., and {Peille}, P.,
  ``{Transition-edge sensor pixel parameter design of the microcalorimeter
  array for the x-ray integral field unit on Athena},'' in [{\em Space
  Telescopes and Instrumentation 2016: Ultraviolet to Gamma
  Ray}{\nolinebreak\hspace{0.1em}]},  {\em Proc. SPIE} {\bf 9905},  99052H
  (July 2016).

\bibitem{Akamatsu2018FDM}
{Akamatsu}, H., {Gottardi}, L., {van der Kurr}, J., {de Vries}, C.~P.,
  {Brujin}, M.~P., {Chervenak}, J.~A., {Kiviranta}, M., {van den Linden},
  A.~J., {Jackson}, B.~D., and {Smith}, S.~S., ``{Frequency domain multiplexed
  readout of TES X-ray microcalorimeters for X-IFU on board of Athena},'' {\em
  Journal Of Low Temp. Phy.}  (Jan. 2018).

\bibitem{Ravera2014DRE}
{Ravera}, L., {Cara}, C., {Ceballos}, M.~T., {Barcons}, X., {Barret}, D.,
  {Cl{\'e}dassou}, R., {Cl{\'e}net}, A., {Cobo}, B., {Doumayrou}, E., {den
  Hartog}, R.~H., {van Leeuwen}, B.-J., {van Loon}, D., {Mas-Hesse}, J.~M.,
  {Pigot}, C., and {Pointecouteau}, E., ``{The DRE: the digital readout
  electronics for ATHENA X-IFU},'' in [{\em Space Telescopes and
  Instrumentation 2014: Ultraviolet to Gamma Ray}{\nolinebreak\hspace{0.1em}]},
   {\em Proc. SPIE} {\bf 9144},  91445T (July 2014).

\bibitem{Moseley1988Opt}
{Moseley}, S.~H., {Kelley}, R.~L., {Schoelkopf}, R.~J., {Szymkowiak}, A.~E.,
  and {McCammon}, D., ``{Advances toward high spectral resolution quantum X-ray
  calorimetry},'' {\em IEEE Transactions on Nuclear Science}~{\bf 35},  59--64
  (Feb. 1988).

\bibitem{McCammon2002Bkg}
{McCammon}, D., {Almy}, R., {Apodaca}, E., {Bergmann Tiest}, W., {Cui}, W.,
  {Deiker}, S., {Galeazzi}, M., {Juda}, M., {Lesser}, A., {Mihara}, T.,
  {Morgenthaler}, J.~P., {Sanders}, W.~T., {Zhang}, J., {Figueroa-Feliciano},
  E., {Kelley}, R.~L., {Moseley}, S.~H., {Mushotzky}, R.~F., {Porter}, F.~S.,
  {Stahle}, C.~K., and {Szymkowiak}, A.~E., ``{A High Spectral Resolution
  Observation of the Soft X-Ray Diffuse Background with Thermal Detectors},''
  {\em Astrophysical Journal}~{\bf 576},  188--203 (Sept. 2002).

\bibitem{Lotti2014Bkg}
{Lotti}, S., {Cea}, D., {Macculi}, C., {Mineo}, T., {Natalucci}, L.,
  {Perinati}, E., {Piro}, L., {Federici}, M., and {Martino}, B., ``{In-orbit
  background of X-ray microcalorimeters and its effects on observations},''
  {\em A\&A}~{\bf 569},  A54 (Sept. 2014).

\bibitem{DeLuca2004Soft}
{De Luca}, A. and {Molendi}, S., ``{The 2-8 keV cosmic X-ray background
  spectrum as observed with XMM-Newton},'' {\em A\&A}~{\bf 419},  837--848
  (June 2004).

\bibitem{Perinati2018Div}
{Perinati}, E. et~al., ``{A magnetic repeller to impact the ATHENA/WFI
  background level: concept and preliminary feasibility study},'' in [{\em
  Space Telescopes and Instrumentation 2018: Ultraviolet to Gamma
  Ray}{\nolinebreak\hspace{0.1em}]},  {\em Proc. SPIE} {\bf 10699} (July 2018).

\bibitem{Lehmer2012AGN}
{Lehmer}, B.~D., {Xue}, Y.~Q., {Brandt}, W.~N., {Alexander}, D.~M., {Bauer},
  F.~E., {Brusa}, M., {Comastri}, A., {Gilli}, R., {Hornschemeier}, A.~E.,
  {Luo}, B., {Paolillo}, M., {Ptak}, A., {Shemmer}, O., {Schneider}, D.~P.,
  {Tozzi}, P., and {Vignali}, C., ``{The 4 Ms Chandra Deep Field-South Number
  Counts Apportioned by Source Class: Pervasive Active Galactic Nuclei and the
  Ascent of Normal Galaxies},'' {\em Astrophysical Journal}~{\bf 752},  46
  (June 2012).

\bibitem{Moretti2003CXB}
{Moretti}, A., {Campana}, S., {Lazzati}, D., and {Tagliaferri}, G., ``{The
  Resolved Fraction of the Cosmic X-Ray Background},'' {\em Astrophysical
  Journal}~{\bf 588},  696--703 (May 2003).

\bibitem{Agostinelli2003GEANT}
Agostinelli, S., Allison, J., Amako, K., Apostolakis, J., Araujo, H., Arce, P.,
  Asai, M., Axen, D., Banerjee, S., Barrand, G., Behner, F., Bellagamba, L.,
  Boudreau, J., Broglia, L., Brunengo, A., Burkhardt, H., Chauvie, S., Chuma,
  J., Chytracek, R., Cooperman, G., Cosmo, G., Degtyarenko, P., Dell'Acqua, A.,
  Depaola, G., Dietrich, D., Enami, R., Feliciello, A., Ferguson, C.,
  Fesefeldt, H., Folger, G., Foppiano, F., Forti, A., Garelli, S., Giani, S.,
  Giannitrapani, R., Gibin, D., Cadenas, J.~G., González, I., Abril, G.~G.,
  Greeniaus, G., Greiner, W., Grichine, V., Grossheim, A., Guatelli, S.,
  Gumplinger, P., Hamatsu, R., Hashimoto, K., Hasui, H., Heikkinen, A., Howard,
  A., Ivanchenko, V., Johnson, A., Jones, F., Kallenbach, J., Kanaya, N.,
  Kawabata, M., Kawabata, Y., Kawaguti, M., Kelner, S., Kent, P., Kimura, A.,
  Kodama, T., Kokoulin, R., Kossov, M., Kurashige, H., Lamanna, E., Lampén,
  T., Lara, V., Lefebure, V., Lei, F., Liendl, M., Lockman, W., Longo, F.,
  Magni, S., Maire, M., Medernach, E., Minamimoto, K., de~Freitas, P.~M.,
  Morita, Y., Murakami, K., Nagamatu, M., Nartallo, R., Nieminen, P.,
  Nishimura, T., Ohtsubo, K., Okamura, M., O'Neale, S., Oohata, Y., Paech, K.,
  Perl, J., Pfeiffer, A., Pia, M., Ranjard, F., Rybin, A., Sadilov, S., Salvo,
  E.~D., Santin, G., Sasaki, T., Savvas, N., Sawada, Y., Scherer, S., Sei, S.,
  Sirotenko, V., Smith, D., Starkov, N., Stoecker, H., Sulkimo, J., Takahata,
  M., Tanaka, S., Tcherniaev, E., Tehrani, E.~S., Tropeano, M., Truscott, P.,
  Uno, H., Urban, L., Urban, P., Verderi, M., Walkden, A., Wander, W., Weber,
  H., Wellisch, J., Wenaus, T., Williams, D., Wright, D., Yamada, T., Yoshida,
  H., and Zschiesche, D., ``Geant4—a simulation toolkit,'' {\em Nuclear
  Instruments and Methods in Physics Research Section A: Accelerators,
  Spectrometers, Detectors and Associated Equipment}~{\bf 506}(3),  250 -- 303
  (2003).

\bibitem{Lotti2018BKG}
{Lotti}, S. et~al., ``{Estimates for the background of the ATHENA X-IFU
  instrument: the Cosmic Rays contribution},'' in [{\em Space Telescopes and
  Instrumentation 2018: Ultraviolet to Gamma Ray}{\nolinebreak\hspace{0.1em}]},
   {\em Proc. SPIE} {\bf 10699} (July 2018).

\bibitem{Macculi2016Cryo}
{Macculi}, C., {Argan}, A., {D'Andrea}, M., {Lotti}, S., {Laurenza}, M.,
  {Piro}, L., {Biasotti}, M., {Corsini}, D., {Gatti}, F., {Torrioli}, G.,
  {Fiorini}, M., {Molendi}, S., {Uslenghi}, M., {Mineo}, T., {Bulgarelli}, A.,
  {Fioretti}, V., and {Cavazzuti}, E., ``{The Cryogenic AntiCoincidence
  detector for ATHENA X-IFU: a program overview},'' in [{\em Space Telescopes
  and Instrumentation 2016: Ultraviolet to Gamma
  Ray}{\nolinebreak\hspace{0.1em}]},  {\em Proc. SPIE} {\bf 9905},  99052K
  (July 2016).

\bibitem{D'Andrea2017CryoAC}
{D'Andrea}, M., {Lotti}, S., {Macculi}, C., {Piro}, L., {Argan}, A., and
  {Gatti}, F., ``{The Cryogenic AntiCoincidence detector for ATHENA X-IFU: a
  scientific assessment of the observational capabilities in the hard X-ray
  band},'' {\em Experimental Astronomy}~{\bf 44},  359--370 (Dec. 2017).

\bibitem{Smith2001APEC}
{Smith}, R.~K., {Brickhouse}, N.~S., {Liedahl}, D.~A., and {Raymond}, J.~C.,
  ``{Collisional Plasma Models with APEC/APED: Emission-Line Diagnostics of
  Hydrogen-like and Helium-like Ions},'' {\em Astrophysical Journal}~{\bf 556},
   L91--L95 (Aug. 2001).

\bibitem{Arnaud1996XSPEC}
{Arnaud}, K.~A., ``{XSPEC: The First Ten Years},'' in [{\em Astronomical Data
  Analysis Software and Systems V}{\nolinebreak\hspace{0.1em}]},  {Jacoby},
  G.~H. and {Barnes}, J., eds., {\em Astronomical Society of the Pacific
  Conference Series} {\bf 101},  17 (1996).

\bibitem{MorrisonWabs}
{Morrison}, R. and {McCammon}, D., ``{Interstellar photoelectric absorption
  cross sections, 0.03-10 keV},'' {\em Astrophysical Journal}~{\bf 270},
  119--122 (July 1983).

\bibitem{Anders1989Solar}
{Anders}, E. and {Grevesse}, N., ``{Abundances of the elements - Meteoritic and
  solar},'' {\em Geochimica et Cosmochimica Acta}~{\bf 53},  197--214 (Jan.
  1989).

\bibitem{Eckert2012Emission}
{Eckert}, D., {Vazza}, F., {Ettori}, S., {Molendi}, S., {Nagai}, D., {Lau},
  E.~T., {Roncarelli}, M., {Rossetti}, M., {Snowden}, S.~L., and {Gastaldello},
  F., ``{The gas distribution in the outer regions of galaxy clusters},'' {\em
  A\&A}~{\bf 541},  A57 (May 2012).

\bibitem{Reiprich2013Outskirts}
{Reiprich}, T.~H., {Basu}, K., {Ettori}, S., {Israel}, H., {Lovisari}, L.,
  {Molendi}, S., {Pointecouteau}, E., and {Roncarelli}, M., ``{Outskirts of
  Galaxy Clusters},'' {\em Space Sci. Rev}~{\bf 177},  195--245 (Aug. 2013).

\bibitem{Arnaud2005Scaling}
{Arnaud}, M., {Pointecouteau}, E., and {Pratt}, G.~W., ``{The structural and
  scaling properties of nearby galaxy clusters. II. The M-T relation},'' {\em
  A\&A}~{\bf 441},  893--903 (Oct. 2005).

\bibitem{Cash1979}
{Cash}, W., ``{Parameter estimation in astronomy through application of the
  likelihood ratio},'' {\em Astrophysical Journal}~{\bf 228},  939--947 (Mar.
  1979).

\bibitem{Molendi2016Background}
{Molendi}, S., {Eckert}, D., {De Grandi}, S., {Ettori}, S., {Gastaldello}, F.,
  {Ghizzardi}, S., {Pratt}, G.~W., and {Rossetti}, M., ``{A critical assessment
  of the metal content of the intracluster medium},'' {\em A\&A}~{\bf 586},
  A32 (Feb. 2016).

\bibitem{Bozzo2016FW}
{Bozzo}, E., {Barbera}, M., {Genolet}, L., {Paltani}, S., {Sordet}, M.,
  {Branduardi-Raymont}, G., {Rauw}, G., {Sciortino}, S., {Barret}, D., and {Den
  Herder}, J.~W., ``{The Filter Wheel and Filters development for the X-IFU
  instrument on-board Athena},'' {\em ArXiv e-prints}  (Sept. 2016).

\end{thebibliography}
\bibliographystyle{spiebib} % makes bibtex use spiebib.bst

\end{document}